\documentclass[10pt, twocolumn]{IEEEtran}

\usepackage{multirow}
\usepackage{booktabs}
\usepackage{longtable}

\usepackage{cite}

\usepackage{graphicx}
\usepackage{amsmath}
\usepackage{amsfonts}
\usepackage{amssymb}
\usepackage{stfloats}
\usepackage{arydshln}

\usepackage{subfigure}
\graphicspath{{figure/}}

\newtheorem{remark}{Remark}

\usepackage{xcolor}
\definecolor{mygreen}{rgb}{0.0, 0.87, 0.13}

\usepackage{amsfonts, amsmath, mathrsfs, amsbsy, amssymb, dsfont, color}
\usepackage{graphicx}

\newcommand\prefixtext[1]{%
  \ifvmode\else\\\@empty\fi
  \noalign{%
    \penalty0%
    \vbox{\mathstrut}%
    \penalty10000%
    \vskip-\baselineskip
    \penalty10000%
    \vbox to 0pt{%
      \normalbaselines
      \ifdim\linewidth=\columnwidth
      \else
        \parshape\@ne
        \@totalleftmargin\linewidth
      \fi
      \vss
      \noindent#1\par}%
      \penalty10000%
      \vskip-\baselineskip}%
      \penalty10000}

\newtheorem{theorem}{Theorem}
\newtheorem{lemma}{Lemma}

\newcommand{\qed}{\nobreak \ifvmode \relax \else
      \ifdim\lastskip<1.5em \hskip-\lastskip
      \hskip1.5em plus0em minus0.5em \fi \nobreak
      \vrule height0.75em width0.5em depth0.25em\fi}

\DeclareMathAlphabet{\mathpzc}{OT1}{pzc}{m}{it}
\newcommand{\comment}[1]{}

\def\({\left(}
\def\){\left)}

\def\bth{\boldsymbol{\theta}}

\def\bmu{\boldsymbol{\mu}}
\def\bnu{\boldsymbol{\nu}}

\def\bup{\boldsymbol{\upsilon}}

\def\bom{\boldsymbol{\omega}}
\def\bGa{\boldsymbol{\Gamma}}
\def\bDe{\boldsymbol{\Delta}}
\def\bTh{\boldsymbol{\Theta}}
\def\bth{\boldsymbol{\theta}}
\def\btau{\boldsymbol{\tau}}
\def\bLa{\boldsymbol{\Lambda}}

\def\bSi{\boldsymbol{\Sigma}}

\def\bOm{\boldsymbol{\Omega}}
\def\tr{\text{tr}}
\def\({\left(}
\def\){\right)}
\def\[{\left[}
\def\]{\right]}
\def\BEq{\begin{eqnarray}}
\def\EEq{\end{eqnarray}}
\def\BE*{\begin{eqnarray*}}
\def\EE*{\end{eqnarray*}}
\def\BA{\begin{array}}
\def\EA{\end{array}}

\def\Nn{\nonumber}
\def\0{\mathbf{0}}
\def\1{\mathbf{1}}
\def\a{\mathbf{a}}
\def\A{\mathbf{A}}

\def\b{\mathbf{b}}
\def\B{\mathbf{B}}

\def\c{\mathbf{c}}
\def\C{\mathbf{C}}

\def\d{\mathbf{d}}
\def\D{\mathbf{D}}

\def\e{\mathbf{e}}
\def\E{\mathbf{E}}

\def\F{\mathbf{F}}

\def\G{\mathbf{G}}

\def\H{\mathbf{H}}
\def\I{\mathbf{I}}

\def\L{\mathbf{L}}

\def\m{\mathbf{m}}

\def\N{\mathbf{N}}
\def\O{\mathbf{O}}

\def\P{\mathbf{P}}

\def\Q{\mathbf{Q}}
\def\r{\mathbf{r}}

\def\R{\mathbf{R}}

\def\S{\mathbf{S}}

\def\T{\mathbf{T}}
\def\u{\mathbf{u}}

\def\v{\mathbf{v}}

\def\w{\mathbf{w}}
\def\W{\mathbf{W}}
\def\x{\mathbf{x}}
\def\X{\mathbf{X}}
\def\y{\mathbf{y}}
\def\Y{\mathbf{Y}}

\def\Z{\mathbf{Z}}

\def\and{\prefixtext{and}}

\def\diag{{\rm diag}}
\def\Diag{{\rm Diag}}

\def\Im{\mathrm{Im}}
\def\Re{\mathrm{Re}}

\title{One-Bit Target Detection in Colocated MIMO Radar with Colored Background Noise}
\author{Yu-Hang Xiao,~\IEEEmembership{Member,~IEEE}, David Ram{\'\i}rez,~\IEEEmembership{Senior Member,~IEEE}, Lei Huang,~\IEEEmembership{Senior Member,~IEEE}, \\Xiao Peng Li,~\IEEEmembership{Member,~IEEE}, and Hing Cheung So,~\IEEEmembership{Fellow,~IEEE}
\thanks{\textcolor{blue}{This work has been submitted to the IEEE for possible publication. Copyright may be transferred without notice, after which this version may
no longer be accessible.}}
\thanks{Y.-H. Xiao, L. Huang and X. P. Li are with the State Key Laboratory of Radio Frequency Heterogeneous Integration (Shenzhen University), Shenzhen University, Shenzhen 518060, China (e-mail: yuhangxiao@szu.edu.cn, lhuang@szu.edu.cn, x.p.li@szu.edu.cn).}
\thanks{D. Ram{\'\i}rez is with the Department of Signal Theory and Communications, Universidad Carlos III de Madrid, Madrid 28911, Spain, and also with the Gregorio Mara{\~n}{\'o}n Health Research Institute, Madrid 28007, Spain (email: david.ramirez@uc3m.es).}
\thanks{H. C. So is with the Department of Electrical Engineering, City University of Hong Kong, Hong Kong, China (e-mail: hcso@ee.cityu.edu.hk).}
}

\begin{document}

\input{epsf}
\date{}

\maketitle
\maketitle
\begin{abstract}

One-bit sampling has emerged as a promising technique in multiple-input multiple-output (MIMO) radar systems due to its ability to significantly reduce data volume, hardware complexity, and power consumption. Nevertheless, current detection methods have not adequately addressed the impact of colored noise, which is frequently encountered in real scenarios. In this paper, we present a novel detection method that accounts for colored noise in MIMO radar systems. Specifically, we derive Rao's test by computing the derivative of the likelihood function with respect to the target reflectivity parameter and the Fisher information matrix, resulting in a detector that takes the form of a weighted matched filter. To ensure the constant false alarm rate (CFAR) property, we also consider noise covariance uncertainty and examine its effect on the probability of false alarm. The detection probability is also studied analytically. Simulation results demonstrate that the proposed detector provides considerable performance gains in the presence of colored noise.
\end{abstract}

\begin{IEEEkeywords}
One-bit analog-to-digital converter (ADC), multiple-input multiple-output (MIMO) radar, Rao's test, target detection.
\end{IEEEkeywords}

\begin{sloppypar}
\section{Introduction}

One-bit radar, characterized by the use of one-bit analog-to-digital converters (ADC), has witnessed significant advancements in radar processing and imaging domains in recent years~\cite{Ren2017,Jin2020TAES,Ameri2019TSP,Foroozmehr2022TAES,Ni2023TSP,Ameri2019}. 
The primary advantages of one-bit sampling include simplified hardware requirements, reduced data volume, and lower power consumption, making it highly suitable for several applications, particularly on small platforms.

Although the sampling process potentially results in information loss, recent studies have demonstrated that this can be effectively mitigated through advanced signal processing techniques. In some instances, these methods can even enhance the overall system performance, for example, through higher sampling rates~\cite{Xiao2022TVT}. Moreover, one-bit radar has proven capable of performing all the functions of traditional high-bit radar, including direction-of-arrival (DOA) estimation~\cite{Shalom2002TAES,Feng2023TVT,Yu2016SPL,Liu2017ICASSP}, range and Doppler estimation~\cite{Shang2020,Shang2021JSTSP,Xi2020TAES,Xi2020TSP}, detection~\cite{Yang2023,Yang2023TAES}, tracking~\cite{Stein2015TSP}, and imaging~\cite{Zhao2019TGRS,Ge2023TAES}. Consequently, one-bit radar is emerging as an important development direction in the radar field, with profound implications for the design and application of future radar systems, particularly in the context of efficient and accurate target detection.

However, most current research on one-bit radar operates under the assumption of white noise, which is often unrealistic in practice~\cite{Aubry2023TSP,Wang2022TSP,Han2021TSP,Liu2015TAES,Liu2018TSP}. This discrepancy arises because, in the presence of colored noise, the likelihood function is described by central/non-central orthant probabilities, which lack closed-form expressions. Although these probabilities can be evaluated numerically using fast algorithms, they pose challenges for detection tasks since most detection criteria, such as the generalized likelihood ratio test (GLRT), are based on likelihood functions. Concretely, the orthant probabilities make it exceedingly difficult to determine the maximum likelihood estimate (MLE) of the target reflectivity parameter, thereby hindering the formulation of the GLRT. In this paper, we demonstrate that the derivative of the orthant probability with respect to the mean value can be expressed as an orthant probability of a lower dimension, which can also be numerically evaluated. Utilizing this property, we construct a Rao’s test~\cite{Besson2023TSP,Shikhaliev2023TSP,Maio2010TSP} to address this problem, circumventing the need to compute the MLE.

It is important to note that Rao’s test does not need the MLE only in scenarios where the likelihood under the null is simple, which implies no unknown parameters~\cite{Kay_detection}. Consequently, our detector's development is based on the assumption that the noise covariance matrix is known. However, in practical applications, this matrix is often estimated from noise-only samples using, for instance, the algorithms described in~\cite{Xiao2023TSP,Eamaz2022,Liu2021,2022AOS}. Hence, our subsequent analysis focuses on the impact of estimation errors on the detector’s performance. Initially, we explore how an increased false alarm rate can result from noise covariance mismatch. Specifically, we examine the detector’s distribution when the actual noise covariance matrix is assumed to be at a specific mismatched value. These results are then averaged, considering the prior distribution of estimation errors as provided in~\cite{Xiao2023TSP}, using an improved Monte Carlo method. Employing this approach enables a more nuanced understanding of how estimation errors, characterized by the prior distribution, might influence the performance of the detector.

Continuing our analysis, we examine the detector's detection performance and simplify the results to a non-central $\chi^2$ distribution. This simplification provides a clearer view of how noise covariance mismatch affects the detection process, particularly by decreasing the non-centrality parameter. This reduction can be interpreted as a decrease in the “distance” between the null and non-null distributions, which provides valuable insights into the impact of covariance mismatch on the overall detection performance.

Finally, our theoretical findings are validated via computer simulations. It is shown that under realistic settings, the performance degradation due to noise covariance mismatch is almost negligible, especially when the number of noise-only samples is sufficient. This is a common scenario in one-bit processing, as the sampling rate can be significantly increased due to the simple structure of one-bit ADCs. Additionally, the simulation results confirm that the theoretical analysis accurately reflects the impact of noise mismatch on the null distribution, which allows us to adjust the threshold to maintain a constant false alarm rate (CFAR).


The contributions of this paper are summarized as follows:
\begin{enumerate}

\item \textbf{Development of Rao's Test for One-Bit Target Detection in Colored Noise}: This paper extends our previous work on a white noise detector, as documented in~\cite{Xiao2022TVT}. To the best of our knowledge, this is the first study in the field of one-bit radar processing that takes into account colored noise. This advancement marks a significant step forward in enhancing the applicability and accuracy of one-bit radar systems.

\item \textbf{Accurate Characterization of Null and Non-Null Distributions}:
We derive accurate expressions for the null and non-null distributions of the detector, which are crucial for accurately predicting the false alarm and detection probabilities. More importantly, they provide a deeper understanding of the detector's behavior, aiding in a more informed and nuanced approach to radar detection challenges.

\item \textbf{Analysis of Noise Covariance Matrix Mismatch Impact}: We conduct an in-depth study of how discrepancies in the noise covariance matrix influence detection performance, including its impact on the null distribution. Our analysis reveals that the noise covariance matrix mismatch leads to an increased false alarm rate, necessitating adaptive threshold adjustments to preserve the CFAR property. Additionally, by approximating the non-null distribution by a non-central $\chi^2$ distribution, we show that performance degradation can be quantified as a decrease in the non-centrality parameter. This insight provides a clear and direct understanding of how noise covariance matrix mismatch affects system performance.

\end{enumerate}

The remainder of this paper is organized as follows: Section~\ref{sec:signal_model} presents the signal model for one-bit detection in colocated MIMO radar under colored noise conditions. Section~\ref{sec:detector} details the derivation of a detector based on Rao's test. The analysis of its null and non-null distributions is conducted in Sections~\ref{sec:null distribution} and \ref{sec:non-null distribution}, respectively, which also study the effect of noise mismatch. Section~\ref{sec:simulations} provides simulation results to corroborate the theoretical calculations. The paper concludes with a summary of the main findings.

\subsection*{Notation}

Throughout this paper, we use boldface uppercase letters for matrices and boldface lowercase letters for column vectors, while lowercase letters denote scalar quantities. The notation $\A\in\mathbb{R}^{p\times q} \ (\mathbb{C}^{p\times q})$ indicates that $\A$ is a $p\times q$ real (complex) matrix. The $(i,j)$th entry of $\A$ is denoted by $\A(i,j)$, and $\a(i)$ refers to the $i$th entry of the vector $\a$. The trace of $\A$ is represented as $\tr(\A)$. The function $\Diag(\A)$ retrieves the diagonal matrix of $\A$, and $\diag(\a)$ produces a diagonal matrix with the elements of $\a$. The superscripts $(\cdot)^{-1}$, $(\cdot)^T$, and $(\cdot)^H$ represent the matrix inverse, transpose, and Hermitian transpose operators, respectively. The operators $\mathbb{E}[a]$ and $\mathbb{V}[a]$ denote the expected value and variance of $a$, respectively, while $\mathbb{C}[a,b]$ is the covariance between $a$ and $b$. The symbol $\sim$ means ``distributed as''. The terms $\chi^2_f$ and $\chi^2_f(\delta^2)$ refer, respectively, to the central and non-central Chi-squared distributions, where $f$ is the number of degrees of freedom (DOFs), and $\delta^2$ is the non-centrality parameter. Finally, the operators $\operatorname{Re}(\cdot)$ and $\operatorname{Im}(\cdot)$ extract the real and imaginary parts of their arguments, $\imath$ denotes the imaginary unit, and $\mathrm{sign}(\cdot)$ indicates the sign of its argument.

\section{Signal Model}
\label{sec:signal_model}

We begin by examining a colocated MIMO radar setup, encompassing $p$ transmit and $m$ receive antennas. The transmit array emits a probing signal of length $n$, $\S \in \mathbb{C}^{p \times n}$, towards a specified angle $\phi$, which is then reflected by a far-field point source. The received signal at the input of the ADCs can be represented as:
\begin{equation}  \label{X}
\mathbf{X}= \begin{bmatrix} \x_1,\cdots,\x_n \end{bmatrix} = \beta \mathbf{a}_{r}(\phi) \mathbf{a}_{t}^{T}(\phi) \mathbf{S}+\mathbf{N},
\end{equation}
where $\beta$ is the target reflectivity, $\mathbf{a}_{t}(\phi)\in\mathbb{C}^{p\times 1}$ and $\mathbf{a}_{r}(\phi)\in\mathbb{C}^{m\times 1}$ are the
transmit and receive steering vectors, respectively.
The term $\N\in \mathbb{C}^{m\times n}$ denotes additive noise that is composed of $n$ independent and identically distributed Gaussian vectors with zero mean and covariance matrix $\bSi_{\N}$.

After one-bit quantization, the signal becomes
\begin{equation}\label{Y}
\Y= \begin{bmatrix} \y_1,\cdots,\y_n \end{bmatrix} = \mathcal{Q}(\X) = \mathrm{sign}(\Re(\X)) + \imath \mathrm{sign}(\Im(\X)),
\end{equation}
where $\mathcal{Q}(\cdot)$ is the complex-valued quantization function.

Our objective is to detect the presence or absence of a target by analyzing the quantized observations $\Y$. Hypothesis $\mathcal{H}_1$ means that the target is present, while hypothesis $\mathcal{H}_0$ implies target absence. For the one bit measurements, the problem of target detection boils down to
\begin{equation}\label{quantized_model}
\begin{array}{l}
\mathcal{H}_{0}: \mathbf{Y}=\mathcal{Q}\left(\N\right), \\
\mathcal{H}_{1}: \mathbf{Y}=\mathcal{Q}\left(\beta \mathbf{W}+\mathbf{N} \right).
\end{array}
\end{equation}
That is, we test $\mathcal{H}_{0}: \beta = 0$ vs. $\mathcal{H}_{0}: \beta \neq 0$. To simplify the derivation, we have include all known parameters into a single term, denoted as $\W = \mathbf{a}_{r}(\phi) \mathbf{a}_{t}^{T}(\phi) \mathbf{S}$.

Unlike the scenario discussed in~\cite{Xiao2022TVT}, this paper delves into a more intricate yet commonly encountered situation where the noise is colored, implying that the covariance matrix
$\bSi_{\N}$ deviates from being a diagonal matrix. Under such circumstances, the likelihood function is given by the central/non-central orthant probability, which is markedly more complex than the $Q$ function required in the case of white noise~\cite{Xiao2022TVT}. Consequently, there is a need to devise a new detector to address the case of colored noise, which we obtain under the following assumptions:
\begin{enumerate}
\item The columns of $\N$ are independently and identically distributed (i.i.d.) with a complex circular Gaussian distribution $\mathcal{CN}(\0,\bSi_{\N})$.
\item The covariance matrix
$\bSi_{\N}$ has been estimated using training data and is known to the receiver.
\item The parameter $\beta$ remains invariant throughout the entire observation period.
\end{enumerate}

\section{Detector Design}
\label{sec:detector}

Since the non-central orthant probability lacks a closed-form expression, standard criteria like the GLRT cannot be applied. Instead, we resort to numerical methods for constructing the detector. In this section, we demonstrate that the derivative of the orthant probability can be represented as a lower-dimensional orthant probability, which is computationally tractable. This insight leads us to formulate Rao's test as a weighted sum of squared derivatives.

We begin by stacking the real and imaginary parts of the received signal $\x_i$ as $\underline{\x}_i=
\[\Re(\x_i)^T,
\Im(\x_i)^T\]^T$
and the quantized sigmal $\y_i$ as $\underline{\y}_i=
\[\Re(\y_i)^T,
\Im(\y_i)^T\]^T$.
Given the circular nature of the noise, the covariance matrix of $\underline{\x}_i$ is
\begin{equation}
{\bSi}_{\underline{\x}}=
\frac{1}{2}\begin{bmatrix}
\Re(\bSi_{\N})&-\Im(\bSi_{\N})\\
\Im(\bSi_{\N})&\Re(\bSi_{\N})
\end{bmatrix}.
\end{equation}
Defining $\w_i = \u_i + \imath \v_i$, where $\w_i$ is the $i$th column of $\W$, and $\beta = a + \imath b$, the mean of $\underline{\x}_i$ is
\begin{equation}
    \mathbb{E}[\underline{\x}_i]=\bup_{i}=
    \begin{bmatrix}
a \u_i- b \v_i\\
a\v_i+b\u_i
\end{bmatrix}.
\end{equation}
Additionally, the orthant probability is defined as
\begin{equation}\label{orthant}
   P(\bmu, {\bSi})=\int_0^{\infty} \cdots \int_0^{\infty} \phi_k(\mathbf{x} ; \bmu, {\bSi}) \mathrm{d} x_1 \ldots \mathrm{d} x_k,
\end{equation}
where $\phi_k(\mathbf{x} ; \bmu, {\bSi})$ is the probability density function (PDF) of a $k$-dimensional Gaussian distribution $\mathcal{N}(\bmu,\bSi)$:
\begin{equation}
    \phi_k(\mathbf{x} ; \bmu, {\bSi})=\frac{1}{(2\pi)^{\frac{k}{2}} \left|\bSi \right|^{\frac{1}{2}}}\exp\(-\frac{1}{2}(\x-\bmu)^T\bSi^{-1}(\x-\bmu)\).
\end{equation}

Based on these ingredients, and similar to \cite{Wu2023onebit}, we can write the likelihood of the $i$th observation as
\begin{equation}
\mathcal{L}(\underline{\y}_i ;\bth) = \Pr\{\underline{\y}_i;\bth\} 
= P\(\bmu_i, \bOm_i\),
\end{equation}
where $\bth=[a,b]^T$ encapsulates the real and imaginary components of the unknown reflectivity $\beta=a+\imath b$, $\Z_i = \diag(\underline{\y}_i)$, and $\succ$ denotes the element-wise greater than operator. The means are $\bmu_i=\Z_i \bnu_i$, with $\bnu_i=\D^{-\frac{1}{2}}\bup_{i}$, where $\D=\Diag({\bSi}_{\underline{\x}})$. The matrix $\C=\D^{-\frac{1}{2}}{\bSi}_{\underline{\x}}\D^{-\frac{1}{2}}$ is the coherence matrix~\cite{Coherence_book}, while $\bOm_i=\Z_i\C\Z_i$ account for sample-specific information. Notably, $\bmu_i$ depends on $\bth$ via $\bup_{i}$, whereas $\bOm_i$ do not depend on $\bth$. Considering the independence between samples, the log-likelihood function under $\mathcal{H}_1$ is obtained as:
\begin{align}
    \mathcal{L}(\Y;\bth) = \sum_{i=1}^n \mathcal{L}(\underline{\y}_i ;\bth) = \sum_{i=1}^n\log P(\bmu_i, \bOm_i),
\end{align}
and by noting that under $\mathcal{H}_0$, $\bth = \bth_0 = [0,0]^T$, the log-likelihood function becomes
\begin{align}
    \mathcal{L}(\Y;\bth_0) = \sum_{i=1}^n\log P(\0_{2m}, \bOm_i),
\end{align}
which only depends on the known covariance matrices.

Since there are no unknown parameters under $\mathcal{H}_0$, we propose to use Rao's test to address this problem, given by
\begin{equation}\label{Rao_definition}
T_{\text{R}}=
\left(\!\left.\frac{\partial \mathcal{L}(\mathbf{Y} ; \boldsymbol{\theta})}{\partial \boldsymbol{\theta}}\right|_{\boldsymbol{\theta}=\boldsymbol{\theta}_{0}}
\!\right)^{T}\!\!\mathbf{F}^{-1}\!\!\left(\!\boldsymbol{\theta}_{0}\right)
\left(\left.\frac{\partial \mathcal{L}(\mathbf{Y} ; \boldsymbol{\theta})}{\partial \boldsymbol{\theta}}\right|_{\boldsymbol{\theta}=\boldsymbol{\theta}_{0}}
\!\right),
\end{equation}
where $\mathbf{F}\left(\!\boldsymbol{\theta}\right)$ is the Fisher information matrix (FIM):
\begin{align}\label{FIM_definition}
\mathbf{F}\left(\!\boldsymbol{\theta}\right)=\mathbb{E}\left[ \frac{\partial\mathcal{L}(\Y;\bth)}{\partial \bth}\frac{\partial\mathcal{L}(\Y;\bth)}{\partial \bth^T}  \right].
\end{align}
Denoting $P_i = P(\bmu_i, \bOm_i)$, the derivatives are
\begin{subequations}\label{derivatives}
\begin{align}
\frac{\partial\mathcal{L}(\Y;\bth)}{\partial a}\!=&\!\sum_{i=1}^{n}\frac{1}{P_i}\frac{\partial P_i}{\partial a}\!=\!\sum_{i=1}^{n}\sum_{j=1}^{2m}\frac{1}{P_i}\frac{\partial P_i}{\partial \bmu_{i}(j)}\frac{\partial \bmu_{i}(j)}{\partial a}\\
\frac{\partial\mathcal{L}(\Y;\bth)}{\partial b}\!=&\!\sum_{i=1}^{n}\frac{1}{P_i}\frac{\partial P_i}{\partial b}\!=\!\sum_{i=1}^{n}\sum_{j=1}^{2m}\frac{1}{P_i}\frac{\partial P_i}{\partial \bmu_{i}(j)}\frac{\partial \bmu_{i}(j)}{\partial b},
\end{align}
\end{subequations}
where the right-most derivatives are
\begin{subequations}\label{derivatives_u_ab}
\begin{align}
    \frac{\partial \bmu_{i}}{\partial a}&=\Z_i\D^{-\frac{1}{2}}\begin{bmatrix}
        \u_i\\
        \v_i
    \end{bmatrix}=\Z_i \a_i\\
    \frac{\partial \bmu_{i}}{\partial b}&=\Z_i\D^{-\frac{1}{2}}\begin{bmatrix}
        -\v_i  \\
         \u_i
    \end{bmatrix}=\Z_i\b_i.
\end{align}
\end{subequations}
To compute the derivative of the orthant probability with respect to the mean values, we introduce the following lemma.
\begin{lemma}\label{lemma1}
For an orthant probability $P(\bmu, {\bSi})$, the derivative with respect to the $j$th element of $\bmu$ is
\begin{equation}\label{orthant_derivatiave_mean}
    \frac{\partial P(\bmu, {\bSi})}{\partial \bmu(j)}=\frac{1}{\sqrt{2\pi\bSi(j,j)}} P(\bom(\bmu,j),\R(\bSi,j)),
\end{equation}
where $\bom(\bmu,j)$ denotes the reduced vector after removing the $j$th element of $\bmu$ and
\begin{equation}
\R(\bSi,j)=\bTh(\bSi,j)-\frac{1}{\bSi(j,j)}\r_j  \r_j^T,
\end{equation}
where $\bTh(\bSi,j)$ denotes the reduced matrix after removing the $j$th row and column of $\bSi$, $\bSi(i,j)$ is the $(i,j)$th element of $\bSi$ and $\r_j$ is the reduced vector after removing the $j$th element of the $j$th column of $\bSi$.
\end{lemma}
\begin{IEEEproof}
    See Appendix \ref{appendix:A}.
\end{IEEEproof}
This lemma has demonstrated that the derivative of a
$k$-dimensional orthant probability can be computed from a
$(k-1)$-dimensional orthant probability. Consequently, it enables us to express:
\begin{align}\label{P_i_derivatiave}
   \left. \frac{\partial P_i}{\partial \bmu_i(j)}\right|_{\bth=\0}=\frac{1}{\sqrt{2\pi}}P(\0_{2m-1}, \R(\bOm_i,j)).
\end{align}
Subsequently, the derivatives in \eqref{derivatives} become:
\begin{equation}
\left. \frac{\partial\mathcal{L}(\Y;\bth)}{\partial \bth}\right|_{\bth=\0} = 
\begin{bmatrix}
    \tr(\A^T \L \P^{-1}) \\ \tr(\B^T \L \P^{-1})
\end{bmatrix},
\end{equation}
where the $2m\times n$ matrices $\A$ and $\B$ are defined as:
\begin{align}\label{A_B_definition}
\A&=[\a_1,\cdots,\a_n], &
\B&=[\b_1,\cdots,\b_n],
\end{align}
and $\P$ is the $n\times n$ diagonal matrix:
\begin{align}
\P=\diag(P(\0_{2m}, \bOm_1), \cdots, P(\0_{2m}, \bOm_n)).
\end{align}
The $2m\times n$ matrix $\L$ is defined as:
\begin{align}
    \L=\[\Z_1 \left.\frac{\partial P_1}{\partial \bmu_{1}}\right|_{\bth=\0},\cdots, \left.\Z_n\frac{\partial P_n}{\partial \bmu_{n}}\right|_{\bth=\0}\],\label{L_definition}
\end{align}
with each element given by:
\begin{multline}
\label{eq:derivative_Pi_mui}
\left.\frac{\partial P_i}{\partial \bmu_{i}}\right|_{\bth=\0} \!= \frac{1}{\sqrt{2\pi}}\left[
    P(\0_{2m-1}, \R(\bOm_i,1)), \right.\\
    \left.\cdots , P(\0_{2m-1}, \R(\bOm_i,2m))
\right]^T\!.
\end{multline}
In addition, Appendix \ref{appendix:B} shows that the FIM is
\begin{equation}\label{FIM}
\F(\bth_0) 
=\upsilon^2\I_2,
\end{equation}
where
\begin{equation}
\upsilon^2 =\tr\left( \bDe_1^T\bDe_1\O^{-1}  \right)=\tr\left( \bDe_2^T\bDe_2\O^{-1}  \right),
\end{equation}
with $\O$ defined in \eqref{eq:O_definition}, and $\bDe_1$ and $\bDe_2$ defined in \eqref{eq:deltas_definition}.

As a result, Rao's test is formulated as:
 \begin{equation}\label{T_Rao}
     T_{\rm R}=\frac{1}{\upsilon^2}\Big[\tr^2(\A^T \L \P^{-1})+\tr^2(\B^T \L \P^{-1})\Big]\mathop{\gtrless}\limits_{\mathcal{H}_0}^{\mathcal{H}_1} \gamma.
 \end{equation}
Combining \eqref{T_Rao} with \eqref{L_definition} and \eqref{A_B_definition}, it becomes clear that the detector can be interpreted as a weighted matched filter, where by the weights are calculated from the partial derivatives in matrix $\L$ and the orthant probabilities in matrix $\P$, which are determined by the elements of the noise covariance matrix.

\begin{remark}
To build the detector, it is necessary to calculate $2^{2m}$ orthant probabilities of dimension $2m$ to construct $\O$ and $\P$. We demonstrate in Appendix \ref{appendix:B} that
these probabilities can be grouped into sets of four with identical values, reducing the required calculations to
$2^{2m-2}$.
Regarding the derivatives, our computations involve
$2mn$ orthant probabilities of dimension
$2m-1$, which may be identical for repeated observations. Taking into account the previously discussed symmetry, the computational load for the derivatives is minimized to
$\min(m2^{2m-1}, 2m\tilde{n})$, assuming there are
$\tilde{n} \, (\leq n)$ distinct observations.
Additionally, these probabilities need to be calculated only once for a specific noise covariance matrix and do not require updates unless the matrix changes. Several efficient algorithms for evaluating orthant probabilities, as documented in~\cite{Koyama2015,Miwa2003}, significantly reduce the computational burden in detector construction, enhancing the feasibility of practical, real-world applications.

\end{remark}

\section{Null distribution}
\label{sec:null distribution}

In this section, we delve into the null distribution of the proposed detector $T_{\rm R}$. Initially, we consider the scenario with perfectly known noise covariance matrix. Subsequently, we analyze the effect on the false alarm rate when the noise covariance matrix is known up to some estimation error. Our study employs an improved Monte Carlo approach, which begins with a distribution for a specific error matrix. We then compute the average effect over the prior distribution of this error matrix via Monte Carlo integration, leading to the null distribution for the scenario of imperfect noise covariance matrix estimation.

\subsection{Distribution with Known Noise Covariance}

Let us start by rewriting the test statistic as

\begin{equation}
T_{\text{R}}=w_1^2+w_2^2,
\end{equation}
where
\begin{align}\label{w1}
    w_1&=\frac{1}{\upsilon}\tr(\A^T \L \P^{-1}), &
    w_2&=\frac{1}{\upsilon}\tr(\B^T \L \P^{-1}).
\end{align}
Since the samples are i.i.d. under $\mathcal{H}_0$, it follows straightforwardly from the central limit theorem that $w_1$ and $w_2$ follow asymptotically ($n \rightarrow \infty$) a $2$-dimensional joint Gaussian distribution. In addition, recalling that
\begin{align}
    \w=[w_1,w_2]^T=
        \frac{1}{\upsilon}\[
   \frac{\partial \mathcal{L}(\mathbf{Y} ; \boldsymbol{\theta})}{\partial a},
   \frac{\partial \mathcal{L}(\mathbf{Y} ; \boldsymbol{\theta})}{\partial b}\]^T,
\end{align}
and taking \eqref{FIM} into account, it is easy to show that
\begin{align}
    \mathbb{E}[\w] &= \0_2, & \mathbb{E}[\w\w^T] &= \I_2,
\end{align}
which allows us to conclude that $T_{\text{R}}$ is asymptotically ($n \rightarrow \infty$) Chi-square distributed with 2 degrees of freedom (DoFs),
\begin{equation}\label{matched_null_distribution}
T_\text{R}\sim \chi^2_2.
\end{equation}
That is, $T_\text{R}$ is exponentially distributed with parameter $1/2$, the probability of false alarm becomes
\begin{equation}\label{eq:pfa}
P_{\rm fa}(\gamma) = \text{Pr}\{T_\text{R}>\gamma\}=\exp(-\gamma/2),
\end{equation}
and the detection threshold can be obtained as
\begin{equation}
\gamma=-2\log(P_{\rm fa}).
\end{equation}

\subsection{Distribution with Estimated Noise Covariance}

In the preceding derivations, the noise covariance matrix was assumed to be perfectly known to the receiver. However, in real problems, it must be estimated beforehand using, for instance, the algorithms in~\cite{Xiao2023TSP,Eamaz2022,Liu2021}. Accordingly, we need to account for the mismatch between the true and estimated noise covariance matrices and assess its impact. In this subsection, the scenario where the estimated covariance matrix, $\bSi_{\underline{\x}}$, and the true covariance matrix, $\bSi'_{\underline{\x}}$, differ is examined, aiming to study the changes of the null distribution. We proceed with two assumptions: 1) the true covariance matrix $\bSi'_{\underline{\x}}$ is close to the estimate $\bSi_{\underline{\x}}$, a condition that can be ensured by a sufficient amount of (training) samples, and 2) the prior distribution of $\bSi'_{\underline{\x}}$ given $\bSi_{\underline{\x}}$ is known, which was obtained in~\cite{Xiao2023TSP}.

We start by analyzing the joint distribution of $w_1$ and $w_2$ for the estimated covariance matrix, which is given by the following theorem.

\begin{theorem}\label{theorem:1}
Under noise covariance mismatch, the distribution of $\w=[w_1,w_2]^T$ can be asymptotically ($n \rightarrow \infty$) approximated by the real Gaussian distribution with zero mean and covariance matrix
\begin{equation}\label{w_covariance_1}
\bSi_w = \mathbb{E}[\w\w^T] = \frac{\upsilon^2_1}{\upsilon^2} \I_2,
\end{equation}
where
\begin{equation}\label{upsilon_1}
\upsilon^2_1 =\tr\left( \bDe_1^T\bDe_1\G  \right)=\tr\left( \bDe_2^T\bDe_2\G  \right),
\end{equation}
with
\begin{equation}
  \G=\diag\(\frac{O_1'}{O_1^2},\cdots,\frac{O_\kappa'}{O_\kappa^2}\),
\end{equation}
and $O'_1,\cdots,O'_\kappa$ are defined in \eqref{eq:orthant_prob_mismatch}.
\end{theorem}

\begin{IEEEproof}
See Appendix \ref{appendix:C}.
\end{IEEEproof}

Using Theorem \ref{theorem:1}, we have
\begin{equation}\label{mismatched_null_distribution}
\frac{{\upsilon^2}}{\upsilon^2_1}T_\text{R}\sim \chi^2_2.
\end{equation}
Therefore,  the
probability of false alarm can be rewritten as
\begin{align}\label{eq:pfa_imperfect}
P_{\rm fa}(\gamma) =\exp\left(-\frac{\upsilon^2}{2\upsilon^2_1}\gamma\right),
\end{align}
and the detection threshold is
\begin{align}\label{thre_imperfect}
\gamma=-\frac{2\upsilon^2_1}{\upsilon^2}\log(P_{\rm fa}).
\end{align}

\subsection{Average over Prior Distribution of Estimation Error}

In this section, we proceed to average the null distribution over the prior distribution of the estimation error. Such a prior distribution can be derived using analytical methods detailed in the literature, such as~\cite{Xiao2023TSP}.
Assuming a known prior PDF for $\bSi'_{\underline{\x}}$, denoted by $f(\bSi'_{\underline{\x}})$, the probability of false alarm can be computed as
\begin{align}
    P_{\rm fa}(\gamma)&=\int\text{Pr}\{T_\text{R}>\gamma|\bSi'_{\underline{\x}}\}f(\bSi'_{\underline{\x}})\mathrm{d} \bSi'_{\underline{\x}}\Nn\\
    &=\int \exp\(-\frac{\upsilon^2}{2\upsilon^2_1}\gamma\)f(\bSi'_{\underline{\x}})\mathrm{d} \bSi'_{\underline{\x}}. \label{prior_pfa}
\end{align}
Given that evaluating $\upsilon^2_1$ for each $\bSi'_{\underline{\x}}$ involves computing $2^{2m-2}$ orthant probabilities, which do not have a closed-form, computing this integral directly is very challenging. To circumvent this, we present an improved Monte Carlo method to approximate \eqref{prior_pfa}. Concretely, we generate a sequence of $K$ covariance matrices from the prior distribution $f(\bSi'_{\underline{\x}})$ and approximate the false alarm probability by averaging the outcomes as
\begin{align}\label{prior_pfa_approximate}
    P_{\rm fa}(\gamma)\approx \frac{1}{K}\sum_{i =1}^{K} \exp\left(-\frac{\upsilon^2}{2\upsilon^2_{1,i}}\gamma\right),
\end{align}
where $\upsilon^2_{1,i} =\tr( \bDe_1^T\bDe_1\G_i) = \tr( \bDe_2^T\bDe_2\G_i)$, with
\begin{equation}
  \G_i = \diag\(\frac{O_{1,i}'}{O_1^2},\cdots,\frac{O_{\kappa,i}'}{O_\kappa^2}\),
\end{equation}
and $O'_{j,i}=P(\0_{2m},\C'_{j,i})$,
with $\C'_{j,i}=\bGa_j\C'_i\bGa_j$, where $\C'_{i}$ is the coherence matrix of the $i$th covariance matrix sample, and $\bGa_j$ is defined in Appendix B.

Achieving an accurate approximation requires a large $K$, thereby increasing computational complexity since computing $\upsilon^2_{1,i}$, for $i = 1, \ldots, K$, necessitates the calculation of $2^{2 m}$ orthant probabilities. However, this process can be optimized by using a Taylor’s expansion around $\C_j=\bGa_j\C\bGa_j$, where $\C$ is the coherence matrix of the estimated noise covariance matrix, $\bSi_{\underline{\x}}$, to approximate the orthant probabilities required for $\upsilon^2_{1,i}$, corresponding to each $\bSi'_{\underline{\x}_{i}}$. This approximation can be formulated as:
\begin{equation}\label{taylor_orthant}
    P(\0_{2m}, \C'_{j,i}) \approx P(\0_{2m}, {\C}_j)+\left.\frac{\partial P(\0_{2m}, \C)}{\partial \c}\right|_{\c=\c_j}(\c'_{j,i}-\c_{j}),
\end{equation}
where
\begin{align}
\c = \mathop{\text{vech}}(\C) = [\C(1,2), \cdots, \C(2m-1,2m)]^T
\end{align}
is the vectorization of the upper triangular part of $\C$, excluding the main diagonal, which stacks the free parameters of $\C$ in a vector. Moreover, $\c_j = \mathop{\text{vech}}(\C_j)$ and $\c'_{j,i} = \mathop{\text{vech}}(\C'_{j,i})$. The partial derivative in this expression can be efficiently computed using the following theorem.
\begin{theorem}\label{theorem:2}
The derivative of the orthant probability $P(\0_{k}, \C)$ with respect to the correlation coefficient $\C(r,s)$, $r < s,$ 
is
\begin{equation}
    \frac{\partial P(\0_{2m}, \C)}{\partial \C(r,s)}=
    \frac{P(\0_{k-2}, \bar{\C})}{2 \pi \sqrt{1 - |\C(r,s)|^2}},
\end{equation}
where $\bar{\C}=[\bTh(\bTh(\C^{-1},r),s-1)]^{-1}$.
\end{theorem}

\begin{IEEEproof}
See Appendix \ref{appendix:D}.
\end{IEEEproof}

By employing this approach, we can avoid computing the orthant probability for each $\bSi'_{\underline{\x}_{i}}$. Instead, we need only compute the derivative and then generate a large number of $\bSi'_{\underline{\x}_{i}}$ samples to obtain a reliable approximation to the null distribution. Let us denote:
\begin{equation}\label{sigma_g_appro}
    \upsilon^2_{1,i}=\upsilon^2+\upsilon^2_{\bigtriangleup,i},
\end{equation}
where
\begin{equation}
\upsilon^2_{\bigtriangleup,i} =\tr\left( \bDe_1^T\bDe_1 (\G_i - \O^{-1})  \right)=\tr\left( \bDe_2^T\bDe_2 (\G_i - \O^{-1})  \right),
\end{equation}
and
\begin{equation}
  \G_i - \O^{-1} =\diag\(\frac{\Delta O_{1,i}}{O_1^2},\cdots,\frac{\Delta O_{\kappa,i}}{O_\kappa^2}\).
\end{equation}
In this expression, $\Delta O_{j,i} = O'_{j,i} - O_{j}$ can be approximated using \eqref{taylor_orthant} as
\begin{equation}
    \Delta O_{j,i}=\left.\frac{\partial P(\0_{2m}, \C)}{\partial \c}\right|_{\c= {\c}_{j}}(\c'_{j,i} - {\c}_{j}).
\end{equation}
Thus, orthant probabilities are computed only once for the derivatives of each
$\C_j$
  and not for every realization of
$\bSi'_{\underline{\x}_{i}}$. Subsequently, substituting \eqref{sigma_g_appro} into \eqref{eq:pfa_imperfect} allows the average probability of false alarm to be estimated as:
\begin{align}\label{pfa_mismatched_app}
P_{\rm fa}(\gamma) \approx \frac{1}{K}\sum_{i =1}^{K}\exp\left(-\frac{\upsilon^2}{2\left(\upsilon^2+\upsilon^2_{\bigtriangleup_i}\right)}\gamma\right).
\end{align}

\section{Non-null distribution}
\label{sec:non-null distribution}

In this section, the non-null distribution of the proposed detector is examined. First, a generalized non-central $\chi^2$ distribution is introduced through the analysis of the joint distribution of $w_1$ and $w_2$. Afterwards, a simplified representation is derived for the low signal-to-noise ratio (SNR) scenario using a Taylor's expansion, yielding a standard non-central $\chi^2$ distribution.

\subsection{Fundamental Result}

Let us start by computing the mean and covariance matrix of $\w$ under $\mathcal{H}_1$, which are presented in the following theorem.

\begin{theorem}\label{theorem:3}
    Under $\mathcal{H}_1$, the mean and covariance matrix of $\w$ are
\begin{align}
\u_w &=\frac{1}{\upsilon} \begin{bmatrix}
\tr(\E_1\Q^T)\\
\tr(\E_2\Q^T)
\end{bmatrix}, &
\bSi_w &=
\begin{bmatrix}
\sigma_{1}^2 & \sigma_{12}\\
\sigma_{12} & \sigma_{2}^2
\end{bmatrix},
\end{align}
where $\E_l =\bDe_l \O^{-1}, l = 1,2,$ and $\Q(i,j) = P({\bGa}_{j}\bnu_i,{\bGa}_{j}\C{\bGa}_{j}), i = 1, \dots, n, j = 1, \ldots, \kappa$.
The proof of this result, along with the definition of the elements of the covariance matrix, are given in Appendix \ref{appendix:E}.
\end{theorem}

Having obtained the mean and covariance matrix of $\w$ under $\mathcal{H}_1$, we now define
\begin{align}
\bSi_w &= \P^T\bLa\P, & \m &=\P\bSi_w^{-\frac{1}{2}}\u_w.
\end{align}
Then, as in~\cite{Xiao2022TVT}, the detector can be rewritten as
\begin{equation} \label{eq:Rao_weightedsum}
T_{\text{R}}=\lambda_1 (\nu_1 + m_1)^2+\lambda_2 (\nu_2 + m_2)^2,
\end{equation}
where $\bLa=\diag(\lambda_1,\lambda_2)$, $\m=[m_1,m_2]^T$, and $\nu_1$, $\nu_2$ are mutually independent standard Gaussian random variables. Thus, the detection probability of $T_{\text{R}}$ is given by a general non-central $\chi^2$ distribution:
\begin{align}\label{non_null}
T_{\text{R}}=\sum_{l=1}^2 \lambda_l\chi^2_1(m_l^2),
\end{align}
which can be numerically evaluated~\cite{Imhof1961}.

It is easily proved that in the case of a mismatched noise covariance matrix, the result is adjusted by substituting $\Q$ with $\Q'$, which is defined as follows:
\begin{equation}\label{non-null_mismatched}
    \Q'(i,j)=P({\bGa}_{j}\bnu_i',{\bGa}_{j}\C'{\bGa}_{j}),
\end{equation}
Here, $\C'$ represents the coherence matrix corresponding to the mismatched covariance matrix ${\bSi}'_{\underline{\x}}$ and $\bnu_i'=\diag({\bSi}'_{\underline{\x}})^{-\frac{1}{2}}\bup_{i}$ characterizes the changes in the mean due to the covariance mismatch.  


\subsection{Low-SNR Approximation}
\label{sec:lowSNR_approximation}

Despite its near-exact nature, the aforementioned approximation requires the computation of the orthant probability for all $n$ samples. Moreover, it does not allow for an insightful comparison between matched and mismatched cases. Thus, in this subsection, a simplified approximation is presented for the low-SNR regime via a Taylor's approximation, which avoids additional orthant probability computations and allows for the aforementioned comparison. The derived result is presented in the following theorem.

\begin{theorem}\label{theorem:4}
In the low-SNR regime, where $|\beta|=\mathcal{O}(n^{-\frac{1}{2}})$, the mean and covariance matrix of vector
$\w$ for the matched noise covariance case are:
\begin{align}\label{H1_moments_low_SNR}
    \u_{w}&=\upsilon
    \begin{bmatrix}
      a \\
      b
    \end{bmatrix}+\mathcal{O}(n^{-\frac{1}{2}}), &
    \bSi_{w}&=\I_2+\mathcal{O}(n^{-\frac{1}{2}}).
\end{align}
Furthermore, for the mismatched noise covariance case, they become
\begin{align}\label{H1_moments_low_SNR_mismatched}
    \u_{w}&=\frac{1}{\upsilon}
    \begin{bmatrix}
      a \varsigma^2_1\\
      b \varsigma^2_2
    \end{bmatrix}+\mathcal{O}(n^{-\frac{1}{2}}), &
    \bSi_{\w}&=\frac{\upsilon_1^2}{\upsilon^2}\I_2+\mathcal{O}(n^{-\frac{1}{2}}),
\end{align}
where $\varsigma^2_l = \tr\left( \bDe_l^T\bDe'_l\O^{-1}  \right), l = 1,2$.
\end{theorem}

\begin{IEEEproof}
See Appendix \ref{appendix:F}.
\end{IEEEproof}
Utilizing this result, it is straightforward to show for the matched case that
\begin{equation}\label{non_null_low_SNR}
T_\text{R}\sim \chi^2_2(\delta^2),
\end{equation}
where $\delta^2=\upsilon^2|\beta|^2$. For the mismatched noise covariance matrix, the approximation yields
\begin{equation}\label{non_null_low_SNR_mismatched}
\frac{\upsilon^2}{\upsilon_1^2}T_\text{R}\sim \chi^2_2(\delta'^2),
\end{equation}
where $\delta'^2  = (a^2\varsigma^4_1+b^2\varsigma^4_2)/{\upsilon_1^2}$.

Combined with \eqref{matched_null_distribution} and \eqref{mismatched_null_distribution}, we can see that the detection power is determined by the non-centrality parameters $\delta^2$ and  $\delta'^2$. A direct comparison shows that the noise covariance mismatch has caused the detection performance to decrease from $\delta^2$ to $\delta'^2$.


\section{Numerical Results}
\label{sec:simulations}

In this section, we conduct numerical simulations to validate our theoretical findings. Initially, we evaluate the accuracy of the derived theoretical null distribution in both matched and mismatched scenarios. Subsequently, we examine the accuracy of the derived theoretical non-null distributions in both scenarios, along with their low-SNR approximations. Finally, we compare the detection performance of the proposed detector, using the receiver operating characteristics (ROC) curve, with the existing one-bit white noise detector proposed in~\cite{Xiao2022TVT}.

We examine a colocated multiple-input multiple-output (MIMO) radar system equipped with a uniform linear array, whose inter-element spacing is half the wavelength. Similar to~\cite{Cui2014TSP,Aldayel2016}, we use an orthogonal linear frequency modulation (LFM) waveform for transmission, directed at angle $\theta$:
\begin{equation}
\mathbf{S}(k,l)=\frac{\exp \left\{\frac{\imath }{n} [2 \pi (l-1) + \pi(l-1)^{2} +(k-1)\sin(\theta)]\right\}}{\sqrt{p}},
\end{equation}
where $k=1,\ldots,p$ and $l=1,\ldots,n$. The DOA $\theta$ is set to
 $\pi/4$, unless specified otherwise. The noise covariance matrix is constructed as
\begin{equation}
    \bSi_{\N}=\alpha \H\H^H+\I_{m}.
\end{equation}
Here, $\H\in \mathbb{C}^{m\times m}$ has i.i.d. elements drawn from a standard complex-valued Gaussian distribution, while $\alpha$ serves as a scaling factor to modulate the correlation coefficients. The signal-to-noise ratio (SNR) is then defined  as
\begin{equation}
\textrm{SNR}=10\log_{10}\!\(\frac{p|\beta|^2}{\tr(\bSi_{\N})}\).
\end{equation}

\subsection{Null Distribution}

\begin{figure}[!t]
\centering{\includegraphics[width=0.8\columnwidth]{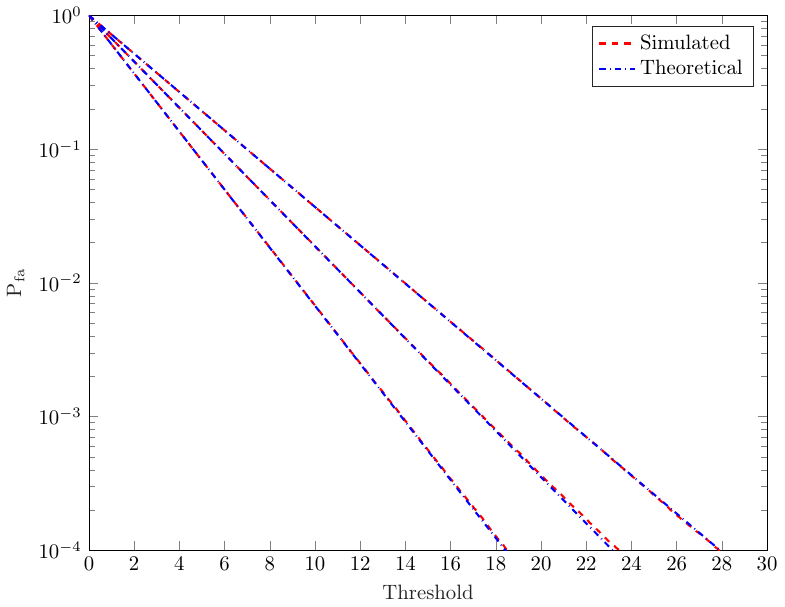}}
\caption{Probability of false alarm versus threshold in an experiment with $m=p=4$, $n=1000$, and $\rho=[0,0.1,0.2]$.}
\label{fig:pfa}
\end{figure}

In this section, we explore two scenarios: one where the noise covariance matrix is precisely known to the receiver, with $\alpha=1$, and another where the actual noise covariance matrix is perturbed, expressed as
\begin{equation}\label{biase_model}
\bSi_{\N}' = \bSi_{{\N}}+\Delta\bSi_{{\N}}.
\end{equation}
The distinct elements of the perturbation $\Delta\bSi_{{\N}}$ are drawn from a normal distribution $\mathcal{N}(\0_{m^2-m},\rho^2\I_{m^2-m})$, with $\rho$ acting as a scaling factor that quantifies the level of error in the covariance matrix. Additionally, to ensure that the generated covariance matrix $\bSi_{\N}'$ remains positive definite, the rare instances where $\bSi_{\N}'$ is not positive definite are excluded from the simulation.

Initially, we examine a scenario where the true covariance matrix is assumed to be a specific value, essentially considering one realization of ~\eqref{biase_model}. Fig.~\ref{fig:pfa} illustrates the probability of false alarms across different thresholds in an experiment characterized by
$m=p=4$ antennas,
$n=1000$, $\alpha=1$ and mismatch levels
$\rho=\{0,0.1,0.2\}$ (curves are ordered from left to right). For the no-mismatch case ($\rho = 0$), we compare the false alarm probability derived from Monte Carlo simulations with the theoretical outcome obtained in~\eqref{eq:pfa}. As depicted in the leftmost curve of the figure, there's an almost perfect alignment between the theoretical predictions and simulation results. This congruence holds true even in the presence of mismatched noise covariance matrices ($\rho \neq 0$), demonstrating the precision of~\eqref{eq:pfa_imperfect}.

\begin{figure}[!t]
\centering{\includegraphics[width=0.8\columnwidth]{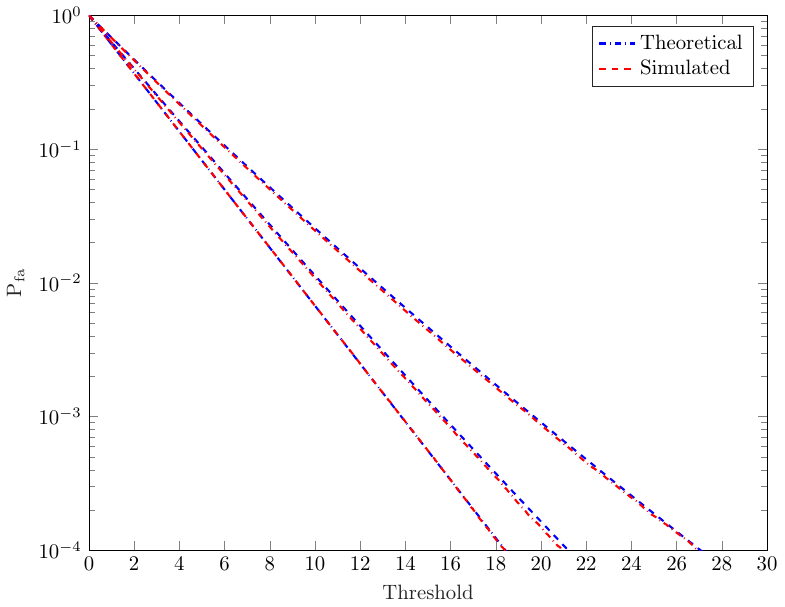}}
\caption{Average probability of false alarm versus threshold in an experiment with $m=p=4$, $n=1000$, and $\rho=[0,0.1,0.2]$.}
\label{fig:average_pfa}
\end{figure}

Subsequently, we evaluate our method's precision for computing the average probability of false alarms.
In this approach, $\bSi_{{\N}}$ is not fixed but generated from $K=1000$ samples according to~\eqref{biase_model}, and the proposed approximation in \eqref{pfa_mismatched_app} is used.
Utilizing the same parameters as in the prior experiment, Fig. \ref{fig:average_pfa} clearly validates the theoretical approximation's precision in~\eqref{pfa_mismatched_app}. Moreover, the advantage of~\eqref{pfa_mismatched_app} lies in its computational efficiency with respect to~\eqref{prior_pfa_approximate}. This efficiency stems from the rapid evaluation of the orthant probability for each sample. In contrast,~\eqref{prior_pfa_approximate} requires computing orthant probabilities for every sample of the noise covariance matrix mismatch, a process that can be markedly time-intensive. As a result, it has been omitted from the simulation.

\subsection{Non-null Distribution}

\begin{figure}[!t]
\begin{minipage}[b]{1\linewidth}
  \centering{\includegraphics[width=0.8\columnwidth]{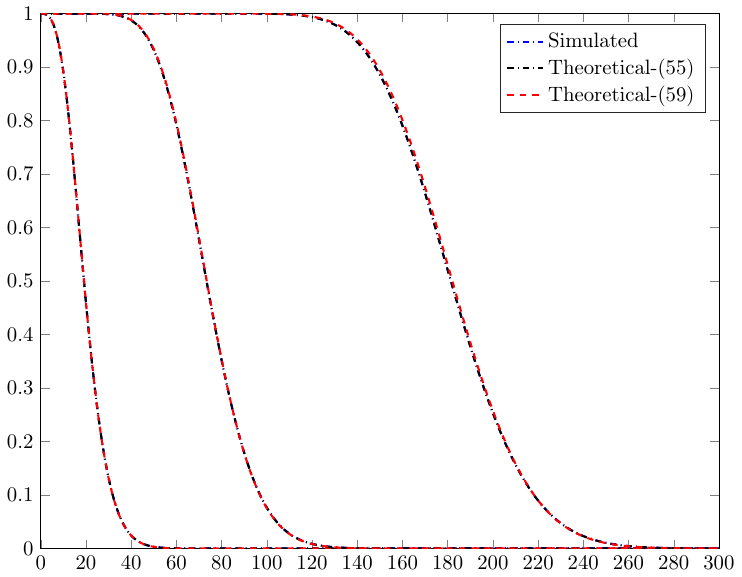}}
\centerline{\small{(a) Matched case}}
\medskip
\end{minipage}
\hfill
\begin{minipage}[b]{1\linewidth}
  \centering{\includegraphics[width=0.8\columnwidth]{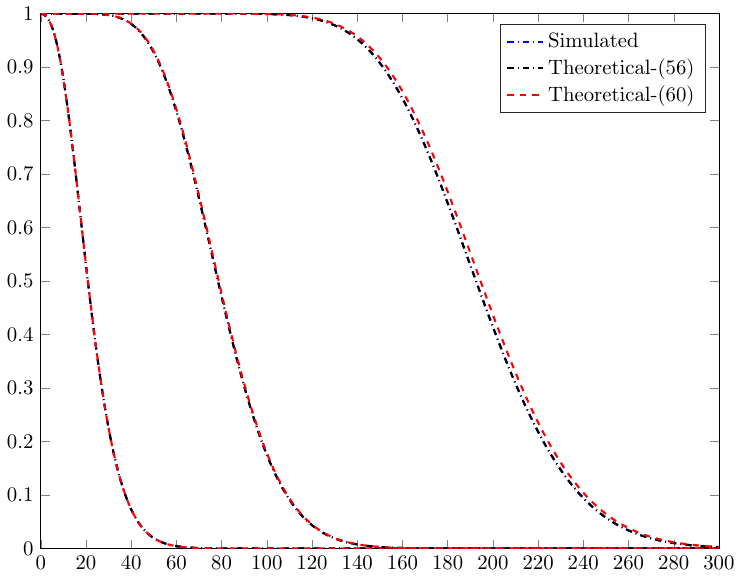}}
\centerline{\small{(b) Mismatched case}}
\medskip
\end{minipage}
\caption{Probability of detection versus threshold in an experiment with $m=p=4$, $n=1000$, and $\text{SNR} = \{-20, -10, -7\}$ dBs.}
\label{fig:pd}
\end{figure}

\begin{table*}[!t]
\begin{center}
\caption{Errors of Different Approximation Methods at Different SNRs}\label{tab:errors}
\begin{tabular}{cccc}
   \toprule[1.5pt]
    & Matched &  & Mismatched \\\midrule[1pt]
  SNR & \!\!\!\!$-20$ dB ~~~~~~~~~$-10$ dB ~~~~~~~~~ $-7$ dB & &
  \!\!\!\!$-20$ dB ~~~~~~~~~$-10$ dB ~~~~~~~~~ $-7$ dB\\\midrule[1pt]
   \!\!Eq.~\eqref{non_null} & $5.58\times 10^{-9}$ ~~ $1.55\times 10^{-8}$ ~~ $4.79\times 10^{-8}$ &~ Eq.~\eqref{non-null_mismatched}&$8.75\times 10^{-9}$ ~~ $2.54\times 10^{-8}$ ~~ $4.93\times 10^{-8}$\\
   \!\!Eq.~\eqref{non_null_low_SNR} & $2.27\times 10^{-8}$ ~~ $8.38\times 10^{-7}$ ~~ $1.78\times 10^{-5}$ &~ Eq.~\eqref{non_null_low_SNR_mismatched}&$2.81\times 10^{-8}$ ~~ $4.31\times 10^{-6}$ ~~ $1.07\times 10^{-4}$\\  \bottomrule[1.5pt]
\end{tabular}
\end{center}
\end{table*}

We now turn our attention to the accuracy of the derived non-null distributions. For the matched scenario, we compare the probability of detection obtained through Monte Carlo simulations with that predicted by~\eqref{non_null} and~\eqref{non_null_low_SNR}. It should be noted that, in practical applications, the detection probability usually ranges in the vicinity of several tenths. Consequently, the figures use a linear scale for the detection probability, in contrast to the logarithmic scale used for the false alarm rate. Fig. \ref{fig:pd}(a) shows these values for an experiment with $m=p=4$, $n=1000$, and $\text{SNR} = \{-20, -10, -7\}$ dBs (from left to right), which shows the accuracy of the theoretical results. Fig. \ref{fig:pd}(b) considers a mismatched scenario, with $\rho = 0.02$, for the same experiment and compares the simulations results with \eqref{non-null_mismatched} and \eqref{non_null_low_SNR_mismatched}. Again, this analysis reveals a high degree of agreement between our theoretical findings and the corresponding simulations.

Additionally, Table \ref{tab:errors} measures the approximation errors for \eqref{non_null} and \eqref{non-null_mismatched}, as well as for the low-SNR approximations, given by \eqref{non_null_low_SNR} and \eqref{non_null_low_SNR_mismatched}. Although the low-SNR approximations achieve worse accuracy, they are notably more straightforward to compute, as discussed in Section \ref{sec:lowSNR_approximation}, offering a practical advantage. 

\subsection{Detection Performance}

In this section, we evaluate the detection performance of the proposed detector and compare it with the white noise one-bit detector derived in~\cite{Xiao2022TVT}. We consider two key scenarios for this evaluation. In the first scenario, the noise covariance matrix is perfectly known to the receiver, and in the second, there is a mismatch in the noise covariance matrix due to estimation errors. For the mismatched scenario, we consider the model described in~\eqref{biase_model}, with $\rho = \{0.01,0.02,0.03\}$. These values are intentionally selected to simulate moderate and significant noise mismatches, mirroring conditions frequently encountered in real-world applications. Our simulations involve a total of $n = 2000$ samples with the SNR fixed at $-25$ dBs.

\begin{figure}[!ht]
\centering{\includegraphics[width=0.8\columnwidth]{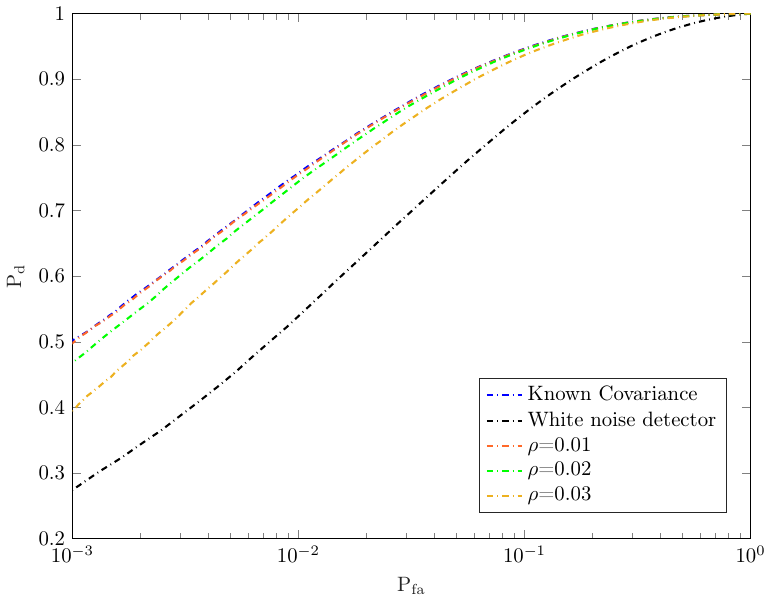}}
\caption{Detection performance under mismatched noise covariance.} \label{fig4}
\end{figure}

As illustrated in Fig.~\ref{fig4}, which shows the ROC curves, the proposed detector shows enhanced detection performance in comparison to the white noise detector across all the considered levels of $\rho$. This finding highlights the critical role of considering the colored nature of the noise. When $\rho = 0.03$, a modest decline in detection performance is observed relative to scenarios with known noise covariance matrix. However, this decline is quantitatively less pronounced than typically expected for such degree of noise mismatch. This effect is attributed to the concurrent rightward shifts in both the null and non-null distributions caused by noise mismatches, as discussed in previous sections, which mitigates the extent of performance degradation. These results imply that at lower mismatch levels ($\rho=\{0.01,0.02\}$), the impact on detection performance is relatively negligible.

Furthermore, according to previous studies on one-bit covariance matrix estimation~\cite{Xiao2023TSP}, estimation errors are often even lower than $\rho=0.02$, thereby supporting the reliability of the proposed detection method in practical scenarios where noise covariance matrix estimation is necessary. Nonetheless, the statistical analysis of noise mismatch impact is still crucial to maintain the CFAR property.

We also explore a more realistic scenario where the noise covariance matrix is estimated using noise-only samples. For this purpose, the observation interval is divided into two windows. The first window collects $n_1$ noise-only samples, which are used for the estimation of the noise covariance matrix using the algorithm presented in~\cite{Xiao2023TSP}. The signal transmission, of length $n_2$, then occurs in the subsequent phase. Since, contrary to the proposed detector, the white noise detector does not use the noise covariance matrix, we exclude the training phase for this detector, dedicating the entire interval to signal transmission. To maintain a fair comparison, the total transmitted power of the signal is kept constant.

\begin{figure}[!ht]
\centering{\includegraphics[width=0.8\columnwidth]{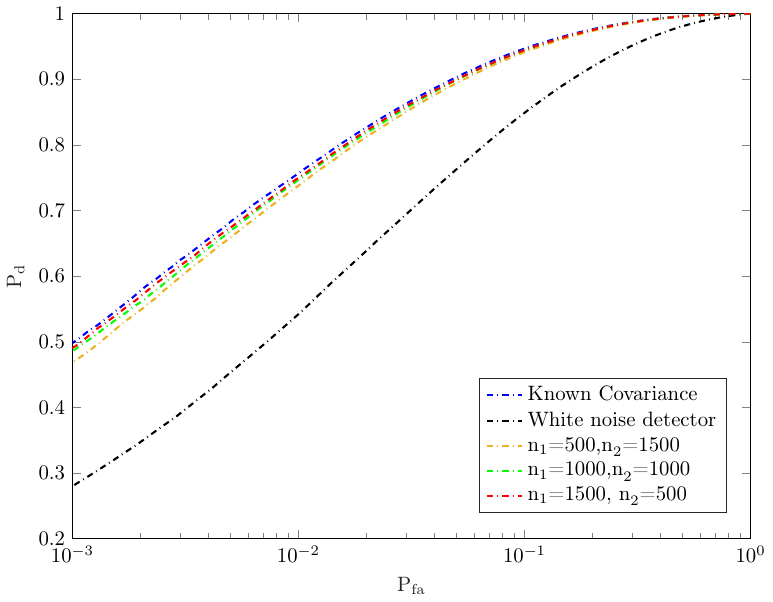}}
\caption{Impact of training sequence length on detection performance.} \label{fig5}
\end{figure}

In the experiments with the proposed detector, we evaluate three configurations for the training sequence: $[n_1, n_2] = [500, 1500]$, $[1000, 1000]$, and $[1500, 500]$. These configurations are chosen to assess the impact of varying lengths of the training sequence on the detection performance. As shown in Fig.~\ref{fig5}, the proposed detector consistently outperforms the white noise detector. Furthermore, the performance curve for all $[n_1, n_2]$ settings closely mirrors that of the scenario with perfectly known noise covariance. This observation confirms that the current estimation approach yields sufficiently precise covariance matrix information, thereby ensuring the robust performance of our detection method. Hence, the effectiveness of the proposed detector in real-world scenarios, where noise covariance estimation is crucial, is underscored by this result.

\section{Conclusion}
\label{sec:conclusions}

In this study, we derived a novel Rao’s test for one-bit target detection in MIMO radar systems operating in colored noise environments, generalizing our prior work~\cite{Xiao2022TVT}. The detector is designed as a weighted matched filter, with weights derived from orthant probabilities tied to noise covariance matrix elements. This approach shows enhanced robustness and significant performance gains in colored noise scenarios compared to the white noise detector~\cite{Xiao2022TVT}. Through comprehensive theoretical analysis, we obtained closed-form approximations for both the null and non-null distributions, enabling accurate calculations of false alarm and detection probabilities. We also  assessed the impact of noise covariance matrix mismatch, highlighting how it increases the false alarm probability and providing the necessary adjustments to maintain the CFAR property. The analysis of the non-null distribution revealed that performance degradation due to covariance mismatch can be quantified by a decrease in the non-centrality parameter of a chi-squared distribution. Simulation results confirmed the effectiveness and practical applicability of the proposed detector in realistic radar detection scenarios.



\appendices

\section{Proof of Lemma \ref{lemma1}}
\label{appendix:A}
We consider $\mu_1$ as an example, as any other derivative can be obtained by a simple permutation of the components of the vector. First, we define the PDF of a zero-mean Gaussian
\begin{equation}
f(x_1,\cdots,x_k)=\phi_k(\mathbf{x} ; \0_k, {\bSi}),
\end{equation}
and rewrite \eqref{orthant} as
\begin{equation}
 P(\bmu,\bSi) = \int_{-\infty}^{\mu_1}\int_{-\infty}^{\mu_2}\cdots\int_{-\infty}^{\mu_{k}}f(x_1,\cdots,x_k)dx_1\cdots  dx_k.
\end{equation}

Taking into account the definition of the partial derivative of $P$ with respect to $\mu_1$, given by
\begin{equation}
 \frac{\partial P}{\partial \mu_1} = \lim_{\Delta \mu_1 \rightarrow 0} \frac{P(\bmu + \Delta \mu_1 \e_1,\bSi) - P(\bmu,\bSi)}{\Delta \mu_1},
\end{equation}
where $\e_1 = [1,0,\cdots,0]^T$, we get
\begin{align}
 \frac{\partial P}{\partial \mu_1}
 &=\lim_{\Delta \mu_1 \rightarrow 0} \frac{\int_{\mu_1}^{\mu_1+\Delta \mu_1}\int_{-\infty}^{\mu_2}\cdots\int_{-\infty}^{\mu_{k}}f(x_1,\cdots,x_k)dx_1\cdots  dx_k}{\Delta \mu_1}\Nn\\
 &=\int_{-\infty}^{\mu_2}\cdots\int_{-\infty}^{\mu_{k}}f(\mu_1,x_2,\cdots,x_k)dx_2\cdots  dx_k.
\end{align}
Now, using Bayes's theorem to decompose the joint PDF, the derivative becomes
\begin{align}
 \frac{\partial P}{\partial \mu_1}
 &=f_{x_1}(\mu_1)\!\int_{-\infty}^{\mu_2}\!\!\!\!\!\cdots\!\!\int_{-\infty}^{\mu_{k}}\!\!f(x_2,\cdots,x_k|x_1=\mu_1)dx_2\cdots  dx_k\Nn\\
&=f_{x_1}(\mu_1)P(\bom(\bmu,1),\R(\bSi,1)),
\end{align}
where $\bom(\bmu,j)$ and $\R(\bSi,j)$ are defined in the lemma. The proof follows from
\begin{equation}
    f_{x_1}(\mu_1)=\frac{1}{\sqrt{2\pi\bSi(1,1)}}.
\end{equation}

\section{Proof of \eqref{FIM} \label{appendix:B}}

Firstly, we arrange all possibilities for $\y$ in ascending order of their binary forms:
\begin{subequations}
\begin{align}
    \btau_1&=[-1,\cdots,-1,-1]^T\\
    \btau_2&=[-1,\cdots,-1,+1]^T\\
    &~~\!\vdots\Nn\\
    \btau_{\kappa}&=[+1,\cdots,+1,+1]^T
\end{align}
\end{subequations}
where $\kappa=2^{2m}$. We define ${\bGa}_{j}=\diag(\btau_j)$, and set
\begin{equation}
    O_j = \Pr\{\underline{\y}=\btau_j|\mathcal{H}_0\}=P(\0,{\bGa}_{j}\C{\bGa}_{j})=P(\0,\C_j),
\end{equation}
and
\begin{equation}\label{d_j_def}
    \d_j = {\bGa}_{j}\left. \frac{\partial P(\bmu,\C_{j})}{\partial \bmu}\right|_{\bmu=\0_{2m}},
\end{equation}
where the partial derivative of $P(\bmu,\C_{j})$ is defined analogously to \eqref{eq:derivative_Pi_mui}.
We first consider the first order statistic:
\begin{equation}
         \mathbb{E}\[ \left.\frac{\partial \mathcal{L}(\underline{\y}_i ;\bth)}{\partial a}\right|_{\bth=\bth_0}\] = \sum_{j=1}^{\kappa}O_j\frac{\a_i^T\d_{j}}{O_j}
         = \a_i^T \bar{\d},
\end{equation}
where $ \bar{\d} = \sum_{j=1}^{\kappa} \d_{j}$.

We proceed to prove that $\bar{\d}=0$. Defining $j^*$ as the index such that $\bGa_{j^*}=-\bGa_{j}$, and given $\bGa_j\C\bGa_j=\bGa_{j^*}\C\bGa_{j^*}$, it is straightforward to demonstrate that $\d_j=-\d_{j^*}$. Since the map between $j$ and $j^*$ is unique, we can conclude that $\bar{\d}=\0$, resulting in:
\begin{equation}
    \label{eq:mean_derivativa_a}
    \mathbb{E}\[ \left.\frac{\partial \mathcal{L}(\underline{\y}_i ;\bth)}{\partial a}\right|_{\bth=\bth_0}\] = 0.
\end{equation}
Similarly, the expectation of the derivative of the log-likelihood function with respect to
$b$ is $0$.

Now, we study the second order statistics. Using \eqref{derivatives} and \eqref{derivatives_u_ab}, we obtain
\begin{equation}\label{FIM11}
         \mathbb{E}\[ \left. \(\frac{\partial \mathcal{L}(\underline{\y}_i ;\bth)}{\partial a}\)^2 \right|_{\bth=\bth_0}\]\!=\!\sum_{j=1}^{\kappa}O_j\frac{(\a_i^T\d_{j})^2}{O_j^2}\!=\!\sum_{j=1}^{\kappa}\frac{(\a_i^T\d_{j})^2}{O_j}.
\end{equation}
Taking into account that observations are independent and \eqref{eq:mean_derivativa_a}, this leads to
\begin{align}\label{sum_a}
    \mathbb{E}\[ \left. \(\frac{\partial \mathcal{L}(\Y ;\bth)}{\partial a}\)^2 \right|_{\bth=\bth_0}\]=\sum_{i=1}^n\sum_{j=1}^{\kappa}\frac{(\a_i^T\d_{j})^2}{O_j}.
\end{align}
Similarly, we have
\begin{align}
    \!\mathbb{E}\[ \left. \(\frac{\partial \mathcal{L}(\Y ;\bth)}{\partial b}\)^2 \right|_{\bth=\bth_0}\]&=\sum_{i=1}^n\sum_{j=1}^{\kappa}\frac{(\b_i^T\d_{j})^2}{O_j},\label{sum_b}\\
    \!\mathbb{E}\[\left. \frac{\partial \mathcal{L}(\Y ;\bth)}{\partial a} \frac{\partial \mathcal{L}(\Y ;\bth)}{\partial b}\right|_{\bth=\bth_0}\]&=\sum_{i=1}^n\sum_{j=1}^{\kappa}\frac{\a_i^T\d_{j}\b_i^T\d_{j}}{O_j}. \label{FIM_cov}
\end{align}
We now explore the ``symmetric'' properties of the orthant probabilities
$\{O_1,\cdots,O_\kappa\}$ and the derivative vectors $\{\d_1,\cdots,\d_\kappa\}$.
Define matrices:
\begin{equation}
   \! \T_1\!=\!\begin{bmatrix}
    \0_m &\! \I_m \\
    -\I_m &\! \0_m
\end{bmatrix}\!,
    \T_2\!=\!\begin{bmatrix}
    \I_m &\! \0_m \\
    \0_m &\! -\I_m
\end{bmatrix}\!,
    \T_3\!=\!\begin{bmatrix}
    \0_m &\! \I_m \\
    \I_m &\! \0_m
\end{bmatrix}
\end{equation}
and let $j_k$ be the integer such that
\begin{equation}
\btau_{j_k}=\T_1^k\btau_j,\quad k=0,1,2,3.
\end{equation}
Given the circular nature of the noise, the coherence matrix remains invariant under the transformations:
\begin{equation}
    \C=\T_1^T\C\T_1=\T_3\T_2\C\T_2\T_3.
\end{equation}
Now, we study the relationships between the orthant probabilities corresponding to $\underline{\y}=\btau_{j}$, $\underline{\y}=\T_2\btau_{j}$, and $\underline{\y}=\T_3\btau_{j}$. It is easy to prove that
\begin{equation}
    \Pr\{\underline{\y}=\T_2\btau_{j};\bth = \bth_0\}=P(\0,\T_2\C_{j}\T_2).
\end{equation}
The transformation $\underline{\y}=\T_3\btau_{j}$ corresponds to a reordering of the elements in $\underline{\y}$. Correspondingly, we can reorder the elements of the coherence matrix, which yields:
\begin{equation}
    \Pr\{\underline{\y}=\T_3\btau_{j};\bth = \bth_0\}=P(\0,\T_3\C_{j}\T_3).
\end{equation}
As a result,
\begin{align}\label{Pj_pT1j}
 O_{j_{1}}&=\Pr\{\underline{\y}=\T_1\btau_{j};\bth = \bth_0\}=\Pr\{\underline{\y}=\T_2\T_3\btau_{j};\bth = \bth_0\}\Nn \\ &=P(\0,\T_3\T_2\C_{j}\T_2\T_3) =P(\0,\C_{j})=O_{j}.
\end{align}
For the derivatives $\d_{j}$ in \eqref{d_j_def}, we use the fact that
\begin{align}
    \C_{j_{1}}=\bGa_{j_{1}}\C\bGa_{j_{1}}=\T_3\bGa_{j}\C\bGa_{j}\T_3=\T_3\C_{j}\T_3,
\end{align}
which correspond to a reordering of the elements in $\C_{j}$. Accordingly, we can reorder the derivative vector in the same manner, leading to
\begin{equation}\label{d_j_relation}
    \left. \frac{\partial P(\bmu,\C_{j_1})}{\partial \bmu}\right|_{\bmu=\0_{2m}}=\T_3\left. \frac{\partial P(\bmu,\C_{j})}{\partial \bmu}\right|_{\bmu=\0_{2m}}.
\end{equation}
In addition, since $\bGa_{j_1}=\T_1\bGa_{j}\T_3$,
by combining \eqref{d_j_def} and \eqref{d_j_relation}, we obtain
\begin{equation}\label{d_def}
    \d_{j_{1}}=\T_{1} \d_{j}.
\end{equation}
Furthermore, it follows from $\eqref{derivatives_u_ab}$ that
\begin{equation}\label{a_b_relation}
\b_i=-\T_1 \a_i.
\end{equation}
Using \eqref{Pj_pT1j}, \eqref{d_def}, \eqref{a_b_relation} and the fact that $\T_1^T\T_1=\I_{2m}$, it is easy to prove that:
\begin{equation}
    \frac{(\a_i^T{\d}_{j})^2}{O_{j}}=\frac{(\b_i^T{\d}_{j_{1}})^2}{O_{j_1}}. 
\end{equation}
Moreover, since $\btau_j=\T_1^4\btau_j$,
we can divide the set $\{1,\cdots, \kappa\}$ into $\kappa/4$ subsets, each with elements $\{j_{k_0},j_{k_1},j_{k_2},j_{k_3}\}$. This division allows us to simplify the summations in \eqref{sum_a} and \eqref{sum_b} as the sum of $\kappa/4$ individual summations. For each subset, the corresponding summation is
\begin{equation}
    \sum_{k=0}^{3}\frac{(\a_i^T{\d}_{j_k})^2}{O_{j_k}}=\sum_{k=0}^{3}\frac{(\b_i^T{\d}_{j_k})^2}{O_{j_k}}.
\end{equation}
Summarizing these subsets yields
\begin{equation}
    \sum_{j=1}^{\kappa}\frac{(\a_i^T{\d}_{j})^2}{O_{j}}=\sum_{j=1}^{\kappa}\frac{(\b_i^T{\d}_{j})^2}{O_{j}}.
\end{equation}
For the covariance term in~\eqref{FIM_cov}, using the symmetries, we can obtain the following relation:
\begin{equation}\label{cov_symmetry}
    \frac{\a_i^T{\d}_{j}\b_i^T{\d}_{j}}{O_j}=-\frac{\a_i^T{\d}_{j_{1}}\b_i^T{\d}_{j_{1}}}{O_{j_{1}}},
\end{equation}
which yields
\begin{equation}
\sum_{k=0}^{3}\frac{\a_i^T{\d}_{j_k}\b_i^T{\d}_{j_k}}{O_{j_k}}=0.
\end{equation}
Consequently, \eqref{FIM_cov} becomes
\begin{align}
\sum_{j=1}^{\kappa}\frac{\a_i^T{\d}_{j}\b_i^T{\d}_{j}}{O_j}=0,
\end{align}
and the FIM simplifies to
\begin{equation}
\F(\bth_0)
=\upsilon^2\I_2,
\end{equation}
where
\begin{equation}
\upsilon^2= \sum_{i=1}^{n}\sum_{j=1}^{\kappa}\frac{(\a_i^T{\d}_{j})^2}{O_{j}}=\sum_{i=1}^{n}\sum_{j=1}^{\kappa}\frac{(\b_i^T{\d}_{j})^2}{O_{j}}.
\end{equation}
Finally, by defining
\begin{equation}
\label{eq:O_definition}
{\O}= \diag\left(O_1,\cdots,O_{\kappa}\right),
\end{equation}
and $\bDe_1$, $\bDe_2\in\mathbb{R}^{n\times \kappa}$ with elements
\begin{align}
\label{eq:deltas_definition}
{\bDe}_{1}(i,j)&=\a_i^T{\d}_{j}, &
{\bDe}_{2}(i,j)&=\b_i^T{\d}_{j},
\end{align}
$\upsilon^2$ can be expressed as
\begin{align}
\upsilon^2 &=\tr\left( \bDe_1^T\bDe_1\O^{-1}  \right)=\tr\left( \bDe_2^T\bDe_2\O^{-1}  \right).
\end{align}
This completes the proof of \eqref{FIM}.

\section{Proof of Theorem \ref{theorem:1} \label{appendix:C}}

It is straightforward to obtain the asymptotic distribution of $\w$ and its mean, so this appendix computes the covariance matrix. Thus, we begin by defining
\begin{equation}
    \label{eq:orthant_prob_mismatch}
    O'_j=P(\0,{\bGa}_{j}\C'{\bGa}_{j}),
\end{equation}
where $\C'$ is the coherence matrix of $\bSi'_{\underline{\x}}$. Following the same argument as \eqref{FIM11} in Appendix \ref{appendix:B}, we have
\begin{equation}
         \mathbb{E}\[\(\frac{\partial \mathcal{L}(\underline{\y}_i ;\bth) }{\partial a}\)^2\]=\sum_{j=1}^{\kappa}O'_j\frac{(\a_i^T\d_{j})^2}{O_j^2}=\sum_{j=1}^{\kappa}(\a_i^T\d_{j})^2\frac{O'_j}{O^2_j}.
\end{equation}
Likewise, we compute
\begin{subequations}
\begin{align}\label{sum_a1}
    \mathbb{E}\[\(\frac{\partial \mathcal{L}(\Y ;\bth)}{\partial a}\)^2\]&=\sum_{i=1}^n\sum_{j=1}^{\kappa}O'_j\frac{(\a_i^T\d_{j})^2}{O_j^2},\\
    \mathbb{E}\[\(\frac{\partial \mathcal{L}(\Y ;\bth)}{\partial b}\)^2\]&=\sum_{i=1}^n\sum_{j=1}^{\kappa}O'_j\frac{(\b_i^T\d_{j})^2}{O_j^2},\\
    \mathbb{E}\[\frac{\partial \mathcal{L}(\Y ;\bth)}{\partial a}\frac{\partial \mathcal{L}(\Y ;\bth)}{\partial b}\]&=\sum_{i=1}^n\sum_{j=1}^{\kappa}O'_j\frac{\a_i^T\d_{j}\b_i^T\d_{j}}{O_j^2}.\label{FIM_cov_mismatched}
\end{align}
\end{subequations}
Once again, by employing the circularity property, it can be shown that
\begin{equation}
O'_{j}=O'_{j_k}, \quad k=0,1,2,3.
\end{equation}
Combining this property with~\eqref{cov_symmetry} and~\eqref{FIM_cov_mismatched}, we have
\begin{equation}
    \mathbb{E}\[\frac{\partial \mathcal{L}(\Y ;\bth)}{\partial a}\frac{\partial \mathcal{L}(\Y ;\bth)}{\partial b}\]=0.
\end{equation}
Analogously, we obtain
\begin{equation}
\mathbb{E}\[\(\frac{\partial \mathcal{L}(\Y ;\bth)}{\partial a}\)^2\]=\mathbb{E}\[\(\frac{\partial \mathcal{L}(\Y ;\bth)}{\partial b}\)^2\]=\upsilon^2_1,
\end{equation}
where $\upsilon^2_1$, i.e., the variance in the mismatched case, is defined in \eqref{upsilon_1}.
Consequently, the covariance matrix of $\w=[w_1,w_2]^T$ is
\begin{equation}
\bSi_w = \frac{\upsilon^2_1}{\upsilon^2} \I_2,
\end{equation}
which completes the proof of Theorem \ref{theorem:1}.

\section{Proof of Theorem \ref{theorem:2} \label{appendix:D}}

The proof of Theorem \ref{theorem:2} is based on Price's Theorem~\cite{Price1958}, which is summarized in the following lemma. For clarity, we present its simplified version.

\begin{lemma}
\label{lem:Price}
Let $\x=[x_1,\cdots,x_k]^T$ be a $k$-dimensional vector following a zero-mean Gaussian distribution with unit variances and coherence matrix $\C$. Consider $k$ nonlinear functions $g_i(x), i=1, 2,\ldots,k,$ and define the $k$th order correlation coefficient of the outputs as
\begin{equation}
R = \mathbb{E}\[\prod_{i=1}^{k} g_i(x_i)\].
\end{equation}
Then, the partial derivative of $R$ with respect to the elements of the coherence matrix $\C$ is given by
\begin{equation}
\frac{\partial R}{ \partial \C(r,s)} = \mathbb{E}\left[  g'_r \left( x_r\right) g'_s \left( x_s\right)\prod_{\substack{i=1\\i\neq r,s}}^k g_i(x_i)\right],
\end{equation}
where $g'_i \left( x_i\right)$ is the derivative of $g_i \left( x_i\right)$ with respect to $x_i$ and $r < s$.
\end{lemma}

To proceed, we define
\begin{equation}
g_i(x)=
\begin{cases}
1, & x\geq 0,\\
0, & x< 0,\\
\end{cases}
\end{equation}
for $i=1,\ldots,k,$ which yields
\begin{equation}
 \mathbb{E}\[\prod_{i=1}^{k} g_i(x_i)\]
= \Pr\{x_1>0,\cdots,x_{k}>0\}
= P(\0_{k},\C).
\end{equation}
In addition, we have
\begin{align}
    g'_r(x_r)g'_s(x_s)=\delta(x_r)\delta(x_s),
\end{align}
where $\delta(\cdot)$ is the Dirac delta function.
Therefore, Lemma \ref{lem:Price} allows us to write the derivative of the orthant probability as
\begin{multline}
\frac{\partial  P(\0_{k},\C)}{ \partial \C(r,s)}
= \mathbb{E}\left[  \delta (x_r) \delta(x_s) \prod_{\substack{i=1\\i\neq r,s}}^k g_i(x_i)\right]\\
=\int_{-\infty}^{\infty} \cdots \int_{-\infty}^{\infty}
\delta(x_r)\delta(x_s)\prod_{\substack{i=1\\i\neq r,s}}^k g_i(x_i)f_\x(\x)d\x.
\end{multline}
Denoting $\bar\x$ as the remaining vector after removing $x_r$ and $x_s$ from $\x$, we can write $f_\x(\x) = f_{\bar{\x}}(\bar{\x}|x_r,x_s) f_{x_r,x_s}(x_r,x_s)$, which yields
\begin{multline}
\frac{\partial  P(\0_{k},\C)}{ \partial \C(r,s)} =f_{x_r,x_s}(0,0)\\
\times\int_{0}^{\infty} \cdots \int_{0}^{\infty} f_{\bar{\x}}(\bar{\x}|0,0)d \bar\x.
\end{multline}
According to the conditional distribution of joint-Gaussian distribution~\cite{Eaton1983}, $\bar\x$ conditioned on $x_r = x_s = 0$ is a zero mean Gaussian vector with covariance matrix
\begin{equation}
\bar{\C}=[\bTh(\bTh(\C^{-1},r),s-1)]^{-1}.
\end{equation}
Therefore, we obtain
\begin{equation}
\int_{0}^{\infty} \cdots \int_{0}^{\infty} f_{\bar{\x}}(\bar{\x}|0,0)d \bar\x = P(\0_{k-2}, \bar{\C}),
\end{equation}
and taking into account that
\begin{align}
f_{x_r,x_s}(0,0) = \frac{1}{2 \pi \sqrt{1 - |\C(r,s)|^2}},
\end{align}
the derivative becomes
\begin{equation}
\frac{\partial P(\0_{k},\C)}{ \partial \C(r,s)} = \frac{P(\0_{k-2}, \bar{\C})}{2 \pi \sqrt{1 - |\C(r,s)|^2}},
\end{equation}
which completes the proof of Theorem \ref{theorem:2}.

\section{Proof of Theorem \ref{theorem:3} \label{appendix:E}}

The definition of the expectation allows us to write
\begin{equation}
    \mathbb{E}\[\frac{\partial \mathcal{L}(\underline{\y}_i ;\bth)}{\partial a}\]=\sum_{j=1}^{\kappa}\Q(i,j)\frac{\a_i^T\d_{j}}{O_j},
\end{equation}
and taking into account the independence of the observations, we get
\begin{align}
    \bmu_w(1)&=\frac{1}{\upsilon}\mathbb{E}\[\frac{\partial \mathcal{L}(\Y ;\bth)}{\partial a}\] =\frac{1}{\upsilon}\sum_{i=1}^n\sum_{j=1}^{\kappa}\Q(i,j)\frac{\a_i^T\d_{j}}{O_j}\Nn\\
    &=\frac{1}{\upsilon}\tr(\E_1\Q^T).
\end{align}
Similarly, we have
\begin{equation}
    \bmu_w(2) = \frac{1}{\upsilon}\mathbb{E}\[\frac{\partial \mathcal{L}(\Y ;\bth)}{\partial b}\]=\frac{1}{\upsilon}\tr(\E_2\Q^T).
\end{equation}
Furthermore, the variance of $w_1$ is
\begin{equation}\label{var1}
   \sigma_1^2 = \frac{1}{\upsilon^2} \mathbb{V}\[\frac{\partial \mathcal{L}(\Y ;\bth)\}}{\partial a}\]
   = \frac{1}{\upsilon^2} \sum_{i=1}^n\mathbb{V}\[\frac{\partial \mathcal{L}(\underline{\y}_i ;\bth)}{\partial a}\],
\end{equation}
where each of the $n$ variances on the right-hand-side of the above expression can be obtained as
\begin{multline}\label{var1_separate}
    \mathbb{V}\[\frac{\partial \mathcal{L}(\underline{\y}_i ;\bth)}{\partial a}\] 
    =\sum_{j=1}^{\kappa} \Q(i,j) \frac{(\a_i^T\d_{j})^2}{O_j^2} \\ -
    \(\sum_{j=1}^{\kappa}\Q(i,j)\frac{\a_i^T\d_{j}}{O_j}\)^2.
\end{multline}
Plugging \eqref{var1_separate} into \eqref{var1} yields
\begin{multline}\label{var1_sum_up}
     \sigma_1^2 = \frac{1}{\upsilon^2} \sum_{i=1}^n\sum_{j=1}^{\kappa}\Q(i,j)\frac{(\a_i^T\d_{j})^2}{O_j^2} \\ - \frac{1}{\upsilon^2} \sum_{i=1}^n\(\sum_{j=1}^{\kappa}\Q(i,j)\frac{\a_i^T\d_{j}}{O_j}\)^2.
\end{multline}
Similarly, we have
\begin{multline}
    \sigma_2^2= \frac{1}{\upsilon^2}\sum_{i=1}^n\sum_{j=1}^{\kappa}\Q(i,j)\frac{(\b_i^T\d_{j})^2}{O_j^2} \\ - \frac{1}{\upsilon^2} \sum_{i=1}^n\(\sum_{j=1}^{\kappa}\Q(i,j)\frac{\b_i^T\d_{j}}{O_j}\)^2,
\end{multline}
and
\begin{multline}
\sigma_{12}= \frac{1}{\upsilon^2}\sum_{i=1}^n\sum_{j=1}^{\kappa}\Q(i,j)\frac{\a_i^T\d_{j}\b_i^T\d_{j}}{O_j^2}\\
- \frac{1}{\upsilon^2}\sum_{i=1}^n\(\sum_{j=1}^{\kappa}\Q(i,j)\frac{\a_i^T\d_{j}}{O_j}\)\times
\(\sum_{j=1}^{\kappa}\Q(i,j)\frac{\b_i^T\d_{j}}{O_j}\).
\end{multline}

\section{Proof of \eqref{H1_moments_low_SNR} and \eqref{H1_moments_low_SNR_mismatched} \label{appendix:F}}

We begin with the case of matched noise covariance. Given that $\beta$ is assumed to be of order $\mathcal{O}(n^{-\frac{1}{2}})$, it can be expressed as
$\beta=\frac{1}{\sqrt{n}}(a_0+\imath b_0)$.
Under such circumstances, a first-order Taylor’s expansion of $\Q(i,j)$ at $[a,b]^T=[0,0]^T$ yields
\begin{multline}
\Q(i,j) =  O_j+\frac{1}{\sqrt{n}}\[a_0 \left.\frac{\partial \Q(i,j)}{\partial a} \right|_{a=0} +b_0 \left.\frac{\partial \Q(i,j)}{\partial b}\right|_{b=0}\]\\
+\mathcal{O}(n^{-1}).
\end{multline}
Taking into account~\eqref{P_i_derivatiave}, the derivatives are:
\begin{subequations}
\begin{align}
   \left.\frac{\partial \Q(i,j)}{\partial a}\right|_{a=0} &= {\a_i^T\d_j}={\bDe_1(i,j)},\\
   \left.\frac{\partial \Q(i,j)}{\partial b}\right|_{b=0} &= {\b_i^T\d_j}={\bDe_2(i,j)},
\end{align}
\end{subequations}
which yields
\begin{equation}
    \label{eq:approx_Q}
    \Q=\1_n \mathbf{o}^T+\frac{1}{\sqrt{n}}\(a_{0}\bDe_1+b_{0}\bDe_2\)+\mathcal{O}(n^{-1}).
\end{equation}
where $\mathbf{o} = [O_1, \cdots, O_{\kappa}]^T$. Since $\E_1=\bDe_1 \O^{-1}$, it can be shown that
\begin{align} \label{eq:temporal_reference}
\u_w(1) &= \frac{1}{\upsilon} \tr(\E_1 \Q^T)\Nn\\
&= \frac{\1_n^T\bDe_1 \1_{\kappa}}{\upsilon} +\frac{1}{\sqrt{n}\upsilon}\left[a_{0} \tr\left(\bDe_1^T \bDe_1 \O^{-1}\right)\right. \Nn\\
    &\phantom{=} \left.+b_{0} \tr\left(\bDe_2^T \bDe_1 \O^{-1}\right)\right]  +  \mathcal{O}(n^{-\frac{1}{2}}).
\end{align}
Using \eqref{eq:deltas_definition}, we have
\begin{equation}
\1_n^T \bDe_1 \1_{\kappa} = \sum_{i = 1}^n \sum_{j = 1}^{\kappa}\a_i^T  \d_j =
\left(\sum_{i = 1}^n \a_i^T \right) \left(\sum_{j = 1}^{\kappa} \d_j \right),
\end{equation}
and recalling that
\begin{equation}
  \sum_{j = 1}^{\kappa} \d_j=\bar{\d}=0,
\end{equation}
we get $\1_n^T\bDe_1 \1_{\kappa}=0$. Moreover, as shown in Appendix~\ref{appendix:B}, $\tr(\bDe_1^T\bDe_2\O^{-1})=0$ and $\tr(\bDe_1^T\bDe_1\O^{-1})=\tr(\bDe_2^T\bDe_2\O^{-1})=\upsilon^2$. Thus, we have
\begin{equation}
\label{eq:u_w_1_low_SNR}
\u_w(1) =  \frac{\upsilon a_0}{\sqrt{n}} + \mathcal{O}(n^{-\frac{1}{2}}),
\end{equation}
and similarly,
\begin{equation}
\label{eq:u_w_2_low_SNR}
    \u_w(2) =  \frac{\upsilon b_0}{\sqrt{n}} + \mathcal{O}(n^{-\frac{1}{2}}).
\end{equation}

For the covariance matrix, using~\eqref{var1_sum_up}, the variance $\sigma^2_1$ is
\begin{align}
\sigma^2_1&=\frac{1}{\upsilon^2}\mathbb{E}\[\(\frac{\partial \mathcal{L}(\Y ;\bth)}{\partial a}\)^2\]-[\u_w(1)]^2\Nn\\
&=1+\frac{1}{\sqrt{n}\upsilon^2}\sum_{i=1}^n\sum_{j=1}^{\kappa}\frac{a_0(\a_i^T\d_{j})^3}{O_j^2}+\mathcal{O}(n^{-1}).
\end{align}
Similarly, $\sigma^2_2$ becomes
\begin{equation}
\sigma^2_2=1+\frac{1}{\sqrt{n}\upsilon^2}\sum_{i=1}^n\sum_{j=1}^{\kappa}\frac{b_0(\b_i^T\d_{j})^3}{O_j^2}+\mathcal{O}(n^{-1}),
\end{equation}
and the covariance $\sigma_{12}$ is
\begin{multline}
{\sigma_{12}}=\frac{1}{\sqrt{n}\upsilon^2}
\sum_{i=1}^n\sum_{j=1}^{\kappa}\frac{\a_i^T\d_{j}\b_i^T\d_{j}}{O_j^2}(a_0\a_i^T\d_{j}+b_0\b_i^T\d_{j})\\
+\mathcal{O}(n^{-1}).
\end{multline}
Given that $\upsilon^2$ is of order $n$, it follows that
\begin{equation}
    \bSi_{w}=\I_2+\mathcal{O}(n^{-\frac{1}{2}}).
\end{equation}

\color{black}
For the mismatched case where the true noise covariance matrix is $\bSi'_{\underline{x}}$, we define:
\begin{subequations}
\begin{align}
    \a'_i&=\Diag(\bSi'_{\underline{x}})^{-\frac{1}{2}}\begin{bmatrix}
        \u_i\\
        \v_i
    \end{bmatrix},\\
    \b'_i&=\Diag(\bSi'_{\underline{x}})^{-\frac{1}{2}}\begin{bmatrix}
        -\v_i  \\
         \u_i
    \end{bmatrix},\\
    \d_j' &= {\bGa}_{j}\left. \frac{\partial P(\bmu,{\bGa}_{j}\C'{\bGa}_{j})}{\partial \bmu}\right|_{\bth=\bth_0}.
\end{align}
\end{subequations}
Mirroring the arguments in the previous subsection, we obtain:
\begin{subequations}
\begin{align}
      \left.\frac{\partial \Q'(i,j)}{\partial a}\right|_{a=0}&=\a'^T_i\d'_j={\bDe'_1(i,j)},\\
      \left.\frac{\partial \Q'(i,j)}{\partial b}\right|_{b=0}&=\b'^T_i\d'_j={\bDe'_2(i,j)}.
\end{align}
\end{subequations}
Thus, the expressions for $u_1$ and $u_2$ are:
\begin{subequations}
\begin{align}
    \u_w(1)&=\frac{a_0}{\sqrt{n}\upsilon}\tr(\bDe_1^T\bDe'_1\O^{-1})\\
    \u_w(2)&=\frac{b_0}{\sqrt{n}\upsilon}\tr(\bDe_2^T\bDe'_2\O^{-1}).
\end{align}
\end{subequations}
As proved in Appendix~\ref{appendix:C}, the covariance matrix under $\mathcal{H}_0$ for the mismatched case  is $\frac{\upsilon_1^2}{\upsilon^2}\I_2$. Following similar arguments, we can conclude that the covariance matrix under low SNR is:
\begin{align}
    \bSi_{w}=\frac{\upsilon_1^2}{\upsilon^2}\I_2+\mathcal{O}(n^{-\frac{1}{2}}).
\end{align}
This completes the proof.

\bibliographystyle{IEEEtran}


\end{sloppypar}

\end{document}